# What can the activation energy tell about the energetics at grain boundaries in polycrystalline organic films?

Lisa S. Walter[a,#,1], Michael Kühn[b,1], Theresa Kammerbauer[c],
James W. Borchert[a], R. Thomas Weitz[a,d,*]

[a] 1st Institute of Physics, Faculty of Physics, Georg-August-University Göttingen, Germany
[b] BASF SE, Pfalzgrafenstr. 1, 67061 Ludwigshafen am Rhein, Germany
[c] Faculty of Physics, Ludwig-Maximilians-University Munich, Germany
[d] International Center for Advanced Studies of Energy Conversion (ICASEC), University of Göttingen, Göttingen, Germany
# current address: Institute of Analytical and Bioanalytical Chemistry, University of Ulm, Germany
[1] These authors contributed equally.
*thomas.weitz@uni-goettingen.de

**ABSTRACT**
Charge-carrier transport at the semiconductor-gate dielectric interface in organic field-effect transistors is critically dependent on the degree of disorder in the typically semi-crystalline semiconductor layer. Charge trapping can occur at the interface as well as in the current-carrying semiconductor layer itself. A detailed and systematic understanding of the role of grain boundaries between crystallites and how to avoid their potentially detrimental effects is still an important focus of research in the organic electronics community. A typical macroscopic measurement technique to extract information about the energetics of the grain boundaries is an activation energy measurement. Here, we compare detailed experiments on the energetic properties of monolayer thin films implemented in organic field-effect transistors, having controlled numbers of grain boundaries within the channel region to kinetic Monte-Carlo simulations of charge-carrier transport to elucidate the influence of grain boundaries on the extracted activation energies. Two important findings are: 1) whereas the energy at the grain boundary does not change with the number of grain boundaries in a thin film, both the measured and simulated activation energy increases with the number of grain boundaries. 2) In simulations where both energy barriers and valleys are present at the grain boundaries there is no systematic relation between the number of grain boundaries and extracted activation energies. We conclude, that a macroscopic measurement of the activation energy can serve as general quality indicator of the thin film, but does not allow microscopic conclusions about the energy landscape of the thin film.

## 1. INTRODUCTION

Organic semiconductors are a core material class at the heart of modern developments of (opto-)electronic devices, including organic photovoltaic devices (OPVs), organic light-emitting diodes (OLEDs) and organic field-effect transistors (OFETs) [1]. In these devices, organic semiconductors are typically implemented in the form of a (poly)crystalline thin film, which typically show energetic disorder in the charge-carrier transport levels at the amorphous regions and grain boundaries between crystallites [2]. Alongside efforts to reduce the overall disorder in organic semiconductors, the detailed characterization of the energetic landscape is critical to increasing the efficiency of these devices. The standard system used to study the relationship between charge-carrier transport and morphological disorder is a (poly)crystalline thin film composed from only one organic semiconductor implemented in an OFET. In this way, many attributes and parameters, such as the component film thicknesses and geometry [3,4], the density of grain boundaries [5,10], the anisotropy within a single crystal [6] and the charge-carrier density [7, 16] can be probed or controlled in concurrence with other factors, such as temperature, to study charge-carrier transport in detail. In particular, due to the pronounced impact of grain boundaries on charge-carrier transport, significant efforts have been dedicated to understanding their detailed energetic properties. This includes especially their tendency to impede charge transport leading to lower mobility [8,9], promote bias-stress related performance degradation [11] and create prime sites for ambient degradation of device performance, e.g. via oxidation [12]. These are generally understood to be due to increased trap state density in the vicinity of grain boundaries. From the theoretical side, the energetics at grain boundaries have been elucidated in great detail [5,10,13,14], with the finding that grain boundaries can introduce higher or lower local activation energies and thereby form energy barriers or valleys, respectively, depending on the neighboring inter-molecular arrangement and the size of the grain boundary [5, 10].

The energetics of films containing grain boundaries can be extracted via temperature and/or charge-carrier density dependent transport measurements, such as the well-established Grünewald method, where the transfer characteristics of an FET is fitted to extract the density and energetic distribution of trap states [15]. Temperature-dependent charge-carrier transport measurements are typically analyzed by Arrhenius-type activation [17]. This allows determination of the density of states (DOS), and also to distinguish between trapping related to the grain boundaries (associated with an exponential tail of the DOS) and trapping within individual grains (associated with Gaussian disorder [10,18,19]). This latter method was implemented in our previous work wherein (poly)crystalline, monolayer-thin films of a perylene diimide derivative (N,N'-di((S)-1-methylpentyl)-1,7(6)-dicyano-perylene-3,4:9,10bis-dicarboximide (PDI1MPCN2)) [20] having different densities of grain boundaries were used as active layers in OFETs. Significantly, the results indicated that the extracted activation energies did not vary systematically between films of different grain-boundary densities, even though their charge-carrier mobilities changed by more than one order of magnitude. This was attributed to the presence of energy barriers, which do not act as trap-sites for charge carriers from which they can be re-excited, but reflect them, and thereby influence the temperature dependence of the thin films in a non-Arrhenius manner [10].

To investigate differences in energy barriers and valleys at grain boundaries in organic semiconductors, simple transport measurements in OFETs may consequently not be sufficient. Instead, Kelvin Probe Force Microscopy (KPFM) has become an established method for the characterization of grain boundaries, since it allows concurrent local probing of the topography and the potential landscape [21,22]. We recently extended this to monolayer PDI1MPCN2 films containing a small density of grain boundaries [23]. Importantly, the use of the monolayer films (as opposed to multilayer films) guarantees that the probed layer is the same layer within which charge-carrier transport takes place. As such, the local charge-carrier density dependence of potential barriers and valleys at the grain boundaries could be identified, which had been suggested also earlier works [22,24]. One of the most significant results from these investigations is that the grain boundaries especially contribute to the conductance at charge-carrier densities below $10^{12}$ cm$^{-2}$, i.e. below the typical working regime of OFETs. In the present work, we show the impacts that the number of grain boundaries within the channel region can have on the activation energy using both temperature-dependent electrical measurements of monolayer PDI1MPCN2 OFETs and corresponding kinetic Monte-Carlo simulations.

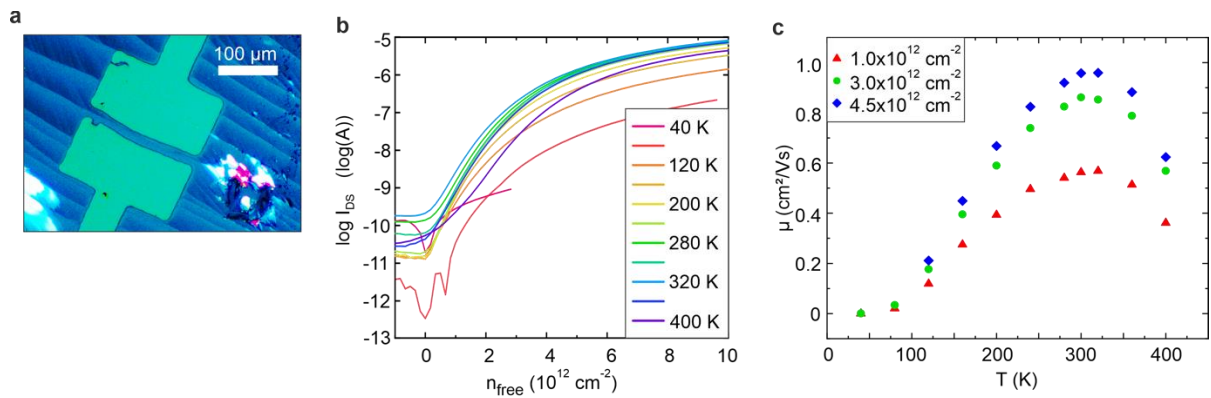

**Figure 1: a).** Polarized optical microscopy image of a monolayer thin PDI1MPCN2 film, grain boundaries are visible as oblique lines in the image. Two metal contacts define the transistor channel that in this particular case does contain a grain boundary. **b)** Temperature dependent transfer curves ($V_{DS}$ = 1 V) as a function of free charge-carrier density $n_{free}$, i.e., normalized to the turn-on voltage $V_{on}$ as described in the text. **c)** Temperature and density dependent charge-carrier mobility.

## 2. RESULTS AND DISCUSSION

Monolayer thin films of the organic semiconductor PDI1MPCN2 were fabricated using previously reported solution processes [20,23] from dimethyl phthalate (DMP) solution onto degenerately-doped Si wavers covered with 30 nm of $Al_2O_3$ formed using atomic layer deposition (ALD). As per reference [23], the process for monolayer formation implemented in this work typically yields energy valleys at the grain boundaries, which are distinguished by higher surface potentials, corresponding to energetically smaller lowest unoccupied molecular orbital (LUMO) levels at grain boundaries than within the surrounding grains. Source and drain contacts for the OFETs were defined using shadow masks with channel lengths of L = 20 μm or L = 200 μm and were applied after semiconductor film formation, and the doped Si substrate was used as the gate (e.g. bottom-gate, top-contact device geometry). The shadow masks were aligned under observation with a polarized optical microscope such that a defined number of grain boundaries were located within the channel (**Fig. 1a**). Transport measurements were performed in vacuum at temperatures ranging from 40 to 400 K using a Lakeshore probe station. The free charge-carrier density $n_{free}$ was controlled via the gate-source voltage $V_{GS}$ and calculated according to the properties of the gate dielectric:

$$n_{free} = \frac{\varepsilon_0 \varepsilon_r}{q d_{ox}}(V_{GS} - V_{on})$$

where $\varepsilon_0$ is the vacuum permittivity, $\varepsilon_r$ and $d_{ox}$ the dielectric constant and the thickness of the insulator ($SiO_2$), respectively, q the elementary charge, and $V_{on}$ the turn-on voltage of the FET [17]. The effective mobility was extracted from the transport measurements in the linear regime using $V_{DS}$ = 1 V (**Fig. 1b**) and used to extract the activation energies ($E_A$) by assuming Arrhenius-like behavior, which were calculated to 27 to 35 meV (Fig. 2) at $n_{free} = 6 \cdot 10^{12}$ cm$^{-2}$. We note that the variation of the activation energy between different samples is larger than the variation of the activation energy between two transistors on the same sample comprising a different amount of grains. We therefore have compared the activation energy only for transistors within the same sample (**Fig. 2**), where one can see that in general the activation energy increases with the number of grain boundaries (by around 10 meV between 0 and 1 grain boundary). This behavior does not show a strong dependence on the free charge-carrier density.

To better understand these experimental results, we performed calculations (described in detail in reference [10]) in thin films, whereby grain boundaries were structurally simulated by attaching neighboring grains which grew from a randomly oriented seeding point. By increasing the number of those seeding points within a specific area, the grain-boundary density of the simulated film was increased. Within these (poly)crystalline films, Levich-Jortner hopping rates were calculated and kinetic Monte-Carlo simulations were used to derive temperature-dependent charge-carrier mobilities (**Fig. 3a**). Hence, we could investigate the impact of the grain-boundary density on the charge-carrier mobility distinct from the impact of the potential difference at the grain boundaries. Performing Arrhenius fits to the data allowed us to extract the activation energy as a function of the number of grains (i.e., grain-boundary density) (**Fig. 3b**). For the films where energy barriers at the grain boundaries have been artificially removed, it can be observed that the extracted activation energy scales with the density of grain boundaries. This result agrees with simulation results obtained in Ref. [5] and with the experimental measurements shown in **Fig 2b**. It is interesting to note that this agreement is good even though the experimental results were obtained at comparably large charge carrier densities ($n_{free} = 6 \cdot 10^{12}$ cm$^{-2}$), while the simulations are performed with single charge carriers traversing the channel. Hence, both, experiments and simulations indicate that the extracted activation energy increases with increasing grain-boundary density, whereby the energy of the trapping sites – both in theory and experiment – does not change. This implies that a measurement of the activation energy cannot help to quantitatively draw conclusions about grain boundary energetics. Moreover, in the case that both valleys and barriers are included in the simulation (**Fig 3b**, black), there is no systematic relation between the density of grain boundaries and the activation energy. Both theory and experiments thus show that caution is needed when trying to draw conclusion about the microscopic behavior of polycrystalline films from a macroscopic measurement of activation energies, i.e., from electrical measurements of OFETs. This observation sheds light on the usability of extracted information form Arrhenius-type fits: The extracted activation energies cannot give specific information about the type (i.e., energy valleys or barriers) nor the magnitude of the true potential difference at a grain boundary. Hence, activation energies should be considered only general indicators of the intrinsic charge-transport properties of the thin films.

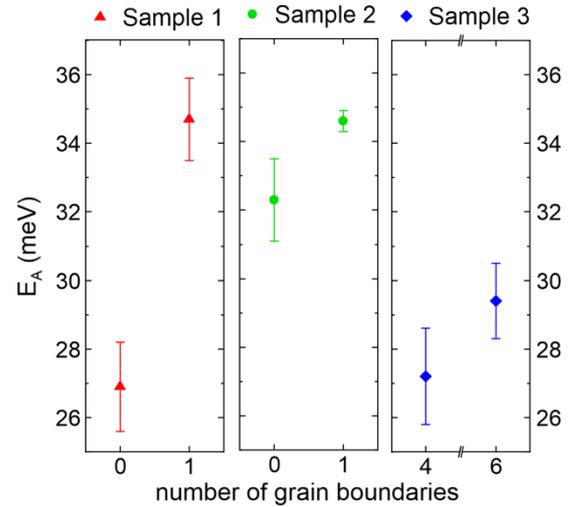

**Figure 2:** Activation energy $E_A$ at a free charge-carrier density of 6x10$^{-12}$ cm$^{-2}$ measured for three different samples (shown in the different colors) plotted against the number of grain boundaries.

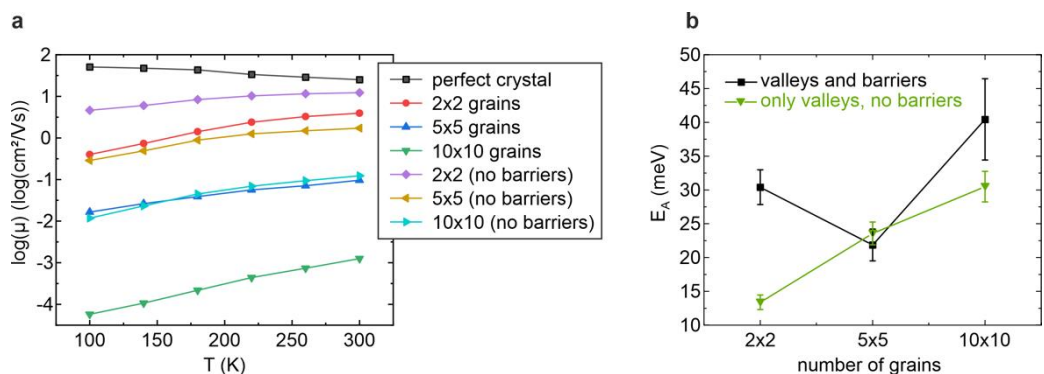

**Figure 3: a)** Simulated temperature-dependent charge-carrier mobility for different number of grain boundary densities. **b)** Extracted activation energy from the data shown in a).

## 3. CONCLUSION

In summary, we have performed a combined theoretical and experimental study about the relationship between the number of grain boundaries and the extracted activation energy in monolayer PDI1MPCN2 thin films. We have shown that, both in experiments on OFETs and through theoretical calculations, the extracted activation energy overall increases with the number of grain boundaries, even though microscopically the grain boundary energetics have not changed. Our work shows that from macroscopic evaluations of the activation energy, one cannot draw conclusions about the true energetics at the grain boundary itself, but the activation energy serves more as a general macroscopic quality indicator of polycrystalline thin films.

**REFERENCES**
[1] Klauk, H. (ed.), "Organic electronics", Wiley-VCH, Weinheim (2008)
[2] Liu, C., Huang, K., Park, W.-T., Li, M., Yang, T., Liu, X., Liang, L., Minari, T. and Noh, Y.-Y., "A unified understanding of charge transport in organic semiconductors: the importance of attenuated delocalization for the carriers" Mater. Horiz. 4(4), 608–618 (2017)
[3] Borchert, J. W., Peng, B., Letzkus, F., Burghartz, J. N., Chan, P. K. L., Zojer, K., Ludwigs, S. and Klauk, H., "Small contact resistance and high-frequency operation of flexible low-voltage inverted coplanar organic transistors" Nat. comm. 10(1), 1119 (2019)
[4] Zhang, F., Di, C., Berdunov, N., Hu, Y., Hu, Y., Gao, X., Meng, Q., Sirringhaus, H. and Zhu, D., "Organic Electronics: Ultrathin Film Organic Transistors: Precise Control of Semiconductor Thickness via Spin-Coating" Adv. Mater. 25(10), 1370 (2013)
[5] Meier, T., Bässler, H. and Köhler, A., "The Impact of Grain Boundaries on Charge Transport in Polycrystalline Organic Field-Effect Transistors" Adv. Opt. Mater. 9(14), 2100115 (2021)
[6] Bittle, E. G., Biacchi, A. J., Fredin, L. A., Herzing, A. A., Allison, T. C., Hight Walker, A. R. and Gundlach, D. J., "Correlating anisotropic mobility and intermolecular phonons in organic semiconductors to investigate transient localization" Comm. Phys. 2(1) (2019)

[7] Podzorov, V., Menard, E., Rogers, J. A. and Gershenson, M. E., "Hall effect in the accumulation layers on the surface of organic semiconductors" Phys. Rev. Lett. 95(22), 226601 (2005)

[8] Chwang, A. B. and Frisbie, C. D., "Temperature and gate voltage dependent transport across a single organic semiconductor grain boundary" J. Appl. Phys. 90(3), 1342–1349 (2001)

[9] Di Carlo, A., Piacenza, F., Bolognesi, A., Stadlober, B. and Maresch, H., "Influence of grain sizes on the mobility of organic thin-film transistors" Appl. Phys. Lett. 86(26), 263501 (2005)

[10] Vladimirov, I., Kühn, M., Geßner, T., May, F. and Weitz, R. T., "Energy barriers at grain boundaries dominate charge carrier transport in an electron-conductive organic semiconductor" Sci. rep. 8 (2018) 14868

[11] Müller, S., Baumann, R.-P., Geßner, T. and Weitz, R. T., "Identification of grain boundaries as degradation site in n-channel organic field-effect transistors determined via conductive atomic force microscopy" Phys. Status Solidi RRL 10(4), 339–345 (2016)

[12] Weitz, R. T., Amsharov, K., Zschieschang, U., Burghard, M., Jansen, M., Kelsch, M., Rhamati, B., van Aken, P. A., Kern, K. and Klauk, H., "The Importance of Grain Boundaries for the Time-Dependent Mobility Degradation in Organic Thin-Film Transistors" Chem. Mater. 21(20), 4949–4954 (2009)

[13] Verlaak, S. and Heremans, P., "Molecular microelectrostatic view on electronic states near pentacene grain boundaries," Phys. Rev. B 75(11) (2007)

[14] Steiner, F., Poelking, C., Niedzialek, D., Andrienko, D., Nelson, J., " Influence of orientation mismatch on charge transport across grain boundaries in tri-isopropylsilylethynyl (TIPS) pentacene thin films" Phys. Chem. Chem. Phys. 19, 10854 (2017)

[15] Grünewald, M., Thomas, P. and Würtz, D., "A Simple Scheme for Evaluating Field Effect Data" phys. stat. sol. (b) 100(2), K139-K143 (1980)

[16] I. Vladimirov, S. Müller, R.-P. Baumann, T. Geßner, Z. Molla, S. Grigorian, A. Köhler, H. Bässler, U. Pietsch, R.T. Weitz, "Dielectric/semiconductor interface limits charge carrier motion at elevated temperatures and large carrier densities in a high-mobility organic semiconductor", Adv. Func. Mater. 29, 1807867 (2019)

[17] Köhler, A. and Bässler, H., "Electronic processes in organic semiconductors. An introduction" Wiley-VCH Verlag GmbH & Co, Weinheim (2015)

[18] Horowitz, G., "Validity of the concept of band edge in organic semiconductors" J. Appl. Phys. 118(11), 115502 (2015)

[19] Zhang, Y., Boer, B. de and Blom, P. W. M., "Trap-free electron transport in poly( p -phenylene vinylene) by deactivation of traps with n -type doping" Phys. Rev. B 81(8) (2010)

[20] Vladimirov, I., Kellermeier, M., Geßner, T., Molla, Z., Grigorian, S., Pietsch, U., Schaffroth, L. S., Kühn, M., May, F. and Weitz, R. T., "High-Mobility, Ultrathin Organic Semiconducting Films Realized by Surface-Mediated Crystallization" Nano Lett. 18(1), 9–14 (2018)

[21] Kelley, T. W., Granstrom, E. and Frisbie, C. D., "Conducting Probe Atomic Force Microscopy. A Characterization Tool for Molecular Electronics" Adv. Mater. 11(3), 261–264 (1999)

[22] Kelley, T. W. and Frisbie, C. D., "Gate Voltage Dependent Resistance of a Single Organic Semiconductor Grain Boundary" J. Phys. Chem. B 105(20), 4538–4540 (2001)

[23] Walter, L. S., Axt, A., Borchert, J. W., Kammerbauer, T., Winterer, F., Lenz, J., Weber, S. A. L. and Weitz, R. T., "Revealing and Controlling Energy Barriers and Valleys at Grain Boundaries in Ultrathin Organic Films" Small 18(34), 2200605 (2022)

[24] Horowitz, G., "Tunneling Current in Polycrystalline Organic Thin-Film Transistors" Adv. Funct. Mater. 13(1), 53–60 (2003)